\documentclass{llncs}

\usepackage[inline]{enumitem}
\usepackage{graphicx}
\usepackage{graphics}
\usepackage[pdf]{graphviz}
\usepackage{amsmath}
\usepackage{mathtools}

\usepackage{amsthm}
\usepackage{float}
\usepackage{amssymb}
\usepackage{stmaryrd}
\SetSymbolFont{stmry}{bold}{U}{stmry}{m}{n}
\usepackage{bm}
\usepackage{dsfont}
\usepackage{scalerel}
\usepackage{xspace}
\usepackage[normalem]{ulem}
\usepackage[dvipsnames]{xcolor}
\usepackage{hyperref}
\hypersetup{  colorlinks = true,
  linkcolor  = blue!75!black,
  citecolor = blue!75!black
}
\usepackage[capitalise,nameinlink]{cleveref}
\usepackage{url}
\usepackage{booktabs}
\usepackage[linesnumbered,lined,commentsnumbered,noend,vlined,ruled]{algorithm2e} \Crefname{algocf}{Algorithm}{Algorithms}

\usepackage{longtable}
\usepackage{multirow}
\usepackage{footnote}

\usepackage{listings}
\usepackage{bold-extra}
\usepackage{relsize}

\usepackage[scaled=0.85]{beramono}
\usepackage[T1]{fontenc}
\usepackage[utf8]{inputenc}
\usepackage{tikz-cd}

\makeatletter
\providecommand{\leftsquigarrow}{  \mathrel{\mathpalette\reflect@squig\relax}}
\newcommand{\reflect@squig}[2]{  \reflectbox{$\m@th#1\rightsquigarrow$}}
\makeatother

\usepackage{tikz}
\usetikzlibrary{arrows,calc,shapes.arrows,shapes.multipart,shapes.geometric, positioning, shadows, automata,fit,decorations.markings,backgrounds}
\pgfdeclarelayer{background}
\pgfdeclarelayer{bg2}
\pgfsetlayers{background,bg2,main} 
\tikzstyle{loc}=[draw=black,rectangle,rounded corners=5pt,inner sep=3pt]
\tikzstyle{myinit}=[loc,double,inner sep=4pt] \tikzstyle{lb}=[pos=1,inner sep=0]
\tikzstyle{mynode}=[fill=red!20,circle,inner sep=0pt,draw]
\tikzstyle{myQ}=[fill=yellow!20, rectangle split, rectangle split parts=3, draw, rectangle split horizontal, rectangle split part align={center, top, bottom}]
\tikzstyle{myarrow}=[fill=red!50, single arrow, draw]
\tikzstyle{myCBQ}=[fill=green!10, draw, minimum height=1.5em, minimum width=3.5em, double copy shadow={shadow xshift=4pt, shadow yshift=4pt, fill=green!10, draw}]

\tikzstyle{bx}=[draw=black,inner sep=2.5mm,thick,minimum height=10mm
                ,rounded corners=3pt,font=\footnotesize]
\tikzstyle{pr}=[circle,draw=black,very thick,fill=white,inner sep=2pt]
\tikzstyle{lbl}=[rotate=0,font=\scriptsize]
\definecolor{dblue}{RGB}{52, 81, 105}
\definecolor{lblue}{RGB}{218, 218, 247}
\tikzstyle{grad}=[draw=dblue,fill=white, postaction={path fading=north, fading angle=45, fill=lblue,draw=dblue}]
\tikzstyle{myhubtkz}=[grad,thick,font={\footnotesize\rm\color{dblue}}]
\tikzstyle{myhublt}=[myhubtkz,minimum height=5mm,inner sep=0.5pt]

\definecolor{myblue}{HTML}{617be3}
\definecolor{myorange}{HTML}{e39c61}
\tikzstyle{var}=[circle,draw=black,font=\bf\ttfamily\footnotesize
                ,minimum height=6mm,fill=myorange!50]
\tikzstyle{fun}=[rectangle,draw=black,inner sep=1.5mm,rounded corners=0.5mm
                ,minimum height=1,font=\bf\ttfamily\footnotesize,fill=myblue!50]
\tikzstyle{source}=[var,regular polygon, regular polygon sides=3,inner sep=2pt]
\tikzstyle{sourcef}=[var,regular polygon, regular polygon sides=3,inner sep=0pt,minimum width=23pt]
\tikzstyle{sourceonce}=[sourcef,ultra thick]
\tikzstyle{sink}=[var,regular polygon, rectangle]
\tikzstyle{arr}=[->,rounded corners=5pt,>=stealth']
\makeatletter
\pgfdeclareshape{document}{
\inheritsavedanchors[from=rectangle] \inheritanchorborder[from=rectangle]
\inheritanchor[from=rectangle]{center}
\inheritanchor[from=rectangle]{north}
\inheritanchor[from=rectangle]{south}
\inheritanchor[from=rectangle]{west}
\inheritanchor[from=rectangle]{east}
\backgroundpath{\southwest \pgf@xa=\pgf@x \pgf@ya=\pgf@y
\northeast \pgf@xb=\pgf@x \pgf@yb=\pgf@y
\pgf@xc=\pgf@xb \advance\pgf@xc by-5pt \pgf@yc=\pgf@yb \advance\pgf@yc by-5pt
\pgfpathmoveto{\pgfpoint{\pgf@xa}{\pgf@ya}}
\pgfpathlineto{\pgfpoint{\pgf@xa}{\pgf@yb}}
\pgfpathlineto{\pgfpoint{\pgf@xc}{\pgf@yb}}
\pgfpathlineto{\pgfpoint{\pgf@xb}{\pgf@yc}}
\pgfpathlineto{\pgfpoint{\pgf@xb}{\pgf@ya}}
\pgfpathclose
\pgfpathmoveto{\pgfpoint{\pgf@xc}{\pgf@yb}}
\pgfpathlineto{\pgfpoint{\pgf@xc}{\pgf@yc}}
\pgfpathlineto{\pgfpoint{\pgf@xb}{\pgf@yc}}
\pgfpathlineto{\pgfpoint{\pgf@xc}{\pgf@yc}}
}
}
\makeatother
\definecolor{darkblue}{HTML}{120136}
\definecolor{middleblue}{HTML}{035aa6}
\definecolor{aqua}{HTML}{40bad5}
\definecolor{newyellow}{HTML}{fcbf1e}
\definecolor{bblue}{HTML}{ececfd}
\definecolor{darkgray}{HTML}{666666}

\pgfdeclarelayer{participant}
\pgfdeclarelayer{loop}
\pgfsetlayers{background,participant,main,loop}

\tikzstyle{sync}=[->]
\tikzstyle{async}=[dashed,->]

\tikzstyle{loop}=[draw=gray,dashed,rounded corners=1mm]
\tikzstyle{lbl}=[]\tikzstyle{participant}=[fill=bblue,rounded corners=0.5mm]
\tikzstyle{event}=[circle,inner sep =0.2]

\newcommand{\wrap}[1]{\begin{tabular}{@{}c@{}}#1\end{tabular}}

\tikzstyle{pr}=[circle,draw=black,thick,fill=myblue!50,inner sep=1pt]

\usepackage[normalem]{ulem}
\usepackage[textsize=scriptsize]{todonotes}
\newcommand{\bgtext}[2]{  \bgroup\markoverwith {\textcolor{#1}{\rule[-0.5ex]{2pt}{11pt}}}\ULon{#2}}

\definecolor{webgreen}{rgb}{0,.5,0}
\definecolor{opcolor}{rgb}{0.4,0,0}
\definecolor{webbrown}{rgb}{.6,0,0}
\definecolor{myred}{rgb}{.7,0,0}
\definecolor{mypurple}{rgb}{.5,0,0.6}
\definecolor{dslcomments}{HTML}{308495}\lstdefinelanguage{reodsl}{
  breaklines=true,
  mathescape=true,
  breakatwhitespace=true,
  basicstyle=\ttfamily\footnotesize,
  keywords=[1]{def,data},
  keywordstyle=[1]\bfseries,
  keywords=[2]{fifoFull,fifo,sync,drain,barrier,lossy,filter,xor,match,build},
  keywordstyle=[2]\color{blue},
    keywordstyle=[3]\color{myred},
  keywords=[4]{List,Either,Int,Bool,Nat,->},
  keywordstyle=[4]\color{mypurple},
  sensitive=true,
  stringstyle=\color{webbrown},
  showstringspaces=false,
  morecomment=[l]{//},
  morecomment=[n]{/*}{*/},
  commentstyle=\color{webgreen},
  morestring=[b]",
  morestring=[b]',
  morestring=[b]""",
  literate=           {<-}{{$\gets$}}2
           {->}{{$\,\to$}}2
           {=>}{{$\,\Rightarrow$}}2
           {<-|}{{$\gets\!\!\!|$}}2
           {<~}{{$\leftsquigarrow$}}2
}

\lstdefinestyle{scalaStyle}{
    basicstyle=\ttfamily\footnotesize,
    keywordstyle=[1]\bfseries\footnotesize,
  keywords=[2]{val,def,override,object},
  keywordstyle=[2]\color{blue}\footnotesize,
  keywords=[3]{view,lts,steps,compareBranchBisim,next,check,SOS},
  keywordstyle=[3]\color{myred}\footnotesize,
  keywords=[4]{Pr,A,St,String,ViewType,Act,State,Set,S1,S2,Unit,Term,List,Seq,
              Array,Parser,Program,Configurator,Show,Code,Mermaid,Text,Semantics,Site},
  keywordstyle=[4]\color{mypurple}\footnotesize,
  stringstyle=\color{green!60!black},
  columns=fullflexible,
  keepspaces,
  literate=           {<-}{{$\gets$}}2
           {->}{{$\,\to$}}2
           {=>}{{\color{mypurple}$\,\Rightarrow$}}2
           {<-|}{{$\gets\!\!\!|$}}2
           {<~}{{$\leftsquigarrow$}}2
}

\newcommand{\code}[1]{\!\!\ensuremath{\text{
\lstinline[language=reodsl,keepspaces]"#1"\xspace
}}}

\newcommand{\scalac}[1]{\!\!\ensuremath{\text{
\lstinline[style=scalastyle]!#1!
}}\!\!}

\newcommand{\bash}[1]{{\small\texttt{\gr{#1}}}\xspace}
\newcommand{\filename}[1]{{\small\texttt{#1}}\xspace}
\lstdefinelanguage{choreo}{
  breaklines=true,
  mathescape=true,
  breakatwhitespace=true,
  basicstyle=\ttfamily\footnotesize,
  keywords=[1]{let,in},
  keywordstyle=[1]\bfseries,
    keywordstyle=[2]\color{blue},
  keywordstyle=[3]\bfseries\color{purple},
  keywordstyle=[4]\sffamily\color{blue},
  keywordstyle=[5]\ttfamily\color{webgreen},
      sensitive=true,
  stringstyle=\color{webbrown},
  showstringspaces=false,
  morecomment=[l]{//},
  morecomment=[n]{/*}{*/},
  commentstyle=\color{webgreen},
  morestring=[b]",
  morestring=[b]',
  morestring=[b]""",
  literate=           {<-}{{$\textcolor{opcolor}{\gets}$}}2
           {->}{{$\,\textcolor{opcolor}{\to}$}}2
           {=>}{{$\,\textcolor{opcolor}{\Rightarrow}$}}2
           {<-|}{{$\textcolor{opcolor}{\gets}\!\!\!|$}}2
                                 {^o}{{$\mathsf{\textcolor{webgreen}{^\circ}}$}}1
           {||}{{$\,\textcolor{opcolor}{\|}\,$}}1
           {*}{{$^{\textcolor{opcolor}{*}}$}}1
           {;}{{$\,$\textcolor{opcolor}{;}$\,$}}1
           {+}{{$\,$\textcolor{opcolor}{+}$\,$}}1
           {(+)}{{$\,$\textcolor{opcolor}{$\boldsymbol{\oplus}$}$\,$}}1
           {[+]}{{$\,$\textcolor{opcolor}{$\boldsymbol{\boxplus}$}$\,$}}1
           {?}{{${\textcolor{opcolor}{?}}$}}1
           {<>}{{${\textcolor{opcolor}{\diamond}}$}}1
           {!}{{${\textcolor{opcolor}{!}}$}}1
           {tau}{{${\textcolor{blue}{\tau}}$}}1
           {0}{{${\mathbf{0}}$}}1
}

\newcommand{\cod}[2][a,b,c,d,e,f,ab,ba,ac,cd,bd,ad]{\!\!\ensuremath{\text{
\lstinline[morekeywords={[4]#1},morekeywords={[5]x,y,w,z}]"#2"\xspace
}}}
}
}
}
\lstdefinelanguage{dot}{
  breaklines=true,
  mathescape=true,
  breakatwhitespace=true,
  basicstyle=\ttfamily\footnotesize,
  keywords=[1]{def,data},
  keywordstyle=[1]\bfseries,
  keywords=[2]{fifoFull,fifo,sync,drain,barrier,lossy,filter,xor,match,build},
  keywordstyle=[2]\color{blue},
    keywordstyle=[3]\color{myred},
  keywords=[4]{List,Either,Int,Bool,Nat,->},
  keywordstyle=[4]\color{mypurple},
  sensitive=true,
  stringstyle=\color{webbrown},
  showstringspaces=false,
  morecomment=[l]{//},
  morecomment=[n]{/*}{*/},
  commentstyle=\color{webgreen},
  morestring=[b]",
  morestring=[b]',
  morestring=[b]"""
                                              }

	\tikzstyle{event} = [inner sep=.5mm, fill=white, draw=black, thick, outer sep=.5mm]
	\tikzstyle{label} = [outer sep=.5mm]
	\tikzstyle{leq} = [-stealth, thick]

\tikzstyle{part}=[
    anchor=base,yshift=5mm,,rounded corners, thick,
    draw={rgb,255: red,206;green,200;blue,221},
    fill={rgb,255: red,236;green,236;blue,253},
    font=\small\sf,minimum height=0mm,minimum width=6mm,inner sep=0mm,baseline,anchor=base]
\tikzstyle{arr}=[->,thick,rounded corners=2pt,above]
\tikzstyle{lbl}=[above,anchor=base,yshift=3pt,font=\scriptsize\sf]
\tikzstyle{inter}=[fill=gray!30,minimum height=8mm]
\tikzstyle{boxname}=[ inner sep=2pt, font=\scriptsize\sf,baseline,
                      draw={rgb,255: red,206;green,200;blue,221},
                      fill={rgb,255: red,236;green,236;blue,253}]
\tikzstyle{box}=[ densely dotted, thick,
                  draw={rgb,255: red,206;green,200;blue,221}]

\tikzstyle{pomchoice}=[fill=bblue, inner sep=3mm 
  , rounded corners=1mm]
\tikzstyle{pomlabel}=[fill=white, rounded corners=1mm,draw=blue!60!black]
\tikzstyle{pomevent}=[circle,inner sep =0.2]
\tikzstyle{pomsetbox}=[draw=black, rounded corners=1mm
  , inner sep=1mm]
\pgfdeclarelayer{choice}
\pgfdeclarelayer{pombg}
\pgfsetlayers{background,pombg,choice,main}

\newcommand{\scala}[2][black]{{\smaller\texttt{\textcolor{#1}{#2}}}\xspace}

\tikzstyle{pl}=[circle,draw=black,inner sep=0pt,minimum width=3mm]

\newcommand{\surl}[1]{{\relscale{0.8}\url{#1}}}
\newcommand{\ssurl}[1]{{\relscale{0.8}\url{#1}}}
\newcommand{\bl}[1]{\textcolor{blue!70!black}{#1}}
\newcommand{\pr}[1]{\textcolor{purple!70!black}{#1}}
\newcommand{\gr}[1]{\textcolor{green!60!black}{#1}}

\newcommand{\ext}[1]{\textcolor{red!60!black}{\textbf{#1}}}

\makeatletter
\relax

\usetikzlibrary{svg.path}

\definecolor{orcidlogocol}{HTML}{A6CE39}
\tikzset{
  orcidlogo/.pic={
    \fill[orcidlogocol] svg{M256,128c0,70.7-57.3,128-128,128C57.3,256,0,198.7,0,128C0,57.3,57.3,0,128,0C198.7,0,256,57.3,256,128z};
    \fill[white] svg{M86.3,186.2H70.9V79.1h15.4v48.4V186.2z}
                 svg{M108.9,79.1h41.6c39.6,0,57,28.3,57,53.6c0,27.5-21.5,53.6-56.8,53.6h-41.8V79.1z M124.3,172.4h24.5c34.9,0,42.9-26.5,42.9-39.7c0-21.5-13.7-39.7-43.7-39.7h-23.7V172.4z}
                 svg{M88.7,56.8c0,5.5-4.5,10.1-10.1,10.1c-5.6,0-10.1-4.6-10.1-10.1c0-5.6,4.5-10.1,10.1-10.1C84.2,46.7,88.7,51.3,88.7,56.8z};
  }
}

\newcommand{\@OrigHeightRecip}{0.00390625}

\newlength{\@curXheight}

\DeclareRobustCommand\orcidlink[1]{%
\texorpdfstring{%
\setlength{\@curXheight}{\fontcharht\font`X}%
\href{https://orcid.org/#1}{\XeTeXLinkBox{\mbox{%
\begin{tikzpicture}[yscale=-\@OrigHeightRecip*\@curXheight,
xscale=\@OrigHeightRecip*\@curXheight,transform shape]
\pic{orcidlogo};
\end{tikzpicture}%
}}}}{}}
\makeatother

\renewcommand{\orcidID}[1]{\orcidlink{#1}}
\newcommand{\CAOS}{\textcolor{purple!70!black}{\textsf{Caos}}\xspace}
\newcommand{\CAOSs}{\textcolor{purple!70!black}{\textsf{Caos}}'\xspace}
\newcommand{\caos}{\textcolor{purple!70!black}{\textsf{Caos}}\xspace}

\begin{document}

\title{
\CAOS: A Reusable Scala Web Animator of Operational Semantics
\\{\normalsize (Extended With Hands-On Tutorial)}
}

\author{
       Jos\'e Proen\c{c}a\inst{1}\orcidID{0000-0003-0971-8919}
  \and Luc Edixhoven\inst{2}\orcidID{0000-0002-6011-9535}
}

\institute{
      CISTER, ISEP, Polytechnic Institute of Porto, Portugal
    \email{pro@isep.ipp.pt}
  \and
  Open University (Heerlen) and CWI (Amsterdam), Netherlands
        \email{led@ou.nl}
}

\maketitle

\begin{abstract}
This tool paper presents \CAOS:
a methodology and a programming framework for \emph{{c}omputer-{a}ided design of structural {o}perational {s}emantics for formal models}.
This framework includes a set of Scala libraries and a workflow to produce visual and interactive diagrams that animate and provide insights over the structure and the semantics of a given abstract model with operational rules. 

\CAOS follows an approach in which theoretical
foundations and a practical tool are built together, as an
alternative to foundations-first design (``tool justifies theory'') or
tool-first design (``foundations justify practice''). The  advantage of
\CAOS is that the tool-under-de\-velop\-ment can immediately be used to
automatically run numerous and sizeable examples in order to identify subtle mistakes,
unexpected outcomes, and unforeseen limitations in the
foundations-under-development, as early as possible.

We share two success stories of \CAOSs methodology and framework in our own teaching and research context, where we analyse a simple while-language and a choreographic language, including their operational rules and the concurrent composition of such rules. We further discuss how others can include \CAOS in their own analysis and Scala tools.
\\[2mm]
\textbf{Demo video:}
\surl{https://zenodo.org/record/7876060} \& \surl{https://youtu.be/Xcfn3zqpubw}
\\[2mm]
\textbf{Hands-on tutorial:}
\cref{sect:lambda}

\end{abstract}

\section{Introduction}

Designing formal methods can be hard. Typical challenges of
formal-methods-related research include identifying and dealing with corner
cases, discovering missing assumptions, finding the right abstraction level,
and---of course---proving theorems (and adequately decomposing them into
lemmas). Curiously, and unlike other scientific disciplines, we find that a large majority
of papers written in our community primarily focuses on research \emph{results}
instead of \emph{methods}. In contrast, this tool paper contributes to \emph{the methodology} of
designing formal methods, with special emphasis on Structural Operational Semantics (SOS): we share our experiences with \emph{\textcolor{purple!70!black}{\underline{c}}omputer-\textcolor{purple!70!black}{\underline{a}}ided design \textcolor{purple!70!black}{\underline{o}}f \textcolor{purple!70!black}{\underline{S}}OS for formal methods} with a set of examples
produced by our toolset \CAOS.
Source code and a compilation of examples can be found at \surl{https://github.com/arcalab/caos}.
 We hope that it may inspire colleagues both to apply our methodology and tools, and to share their own methodology-related experiences to our community's benefit.

In a nutshell, in \CAOS, theoretical foundations and a practical tool are built
together side-by-side, from the start, as an alternative to the more typical
\emph{foun\-dations-first design} (``tool justifies theory'') or \emph{tool-first
design} (``foundations justify practice''). The main advantage of \CAOS is that
the tool-under-development can immediately be used to automatically run numerous
and sizeable examples in order
to identify subtle mistakes, unexpected outcomes, and/or
unforeseen limitations in the foundations-under-development, as early as
possible. Essentially, the primacy of automated examples in \CAOS is similar to
the primacy of automated (unit-)tests in modern agile software engineering.
This need for validation and supporting tools in formal methods has been acknowledged, 
e.g., by Garavel et al.\ in a recent survey over formal methods in critical systems~\cite{DBLP:conf/fmics/GaravelBP20}.

\begin{figure}[tbh]
  \centering
  \includegraphics[width=0.8\textwidth]{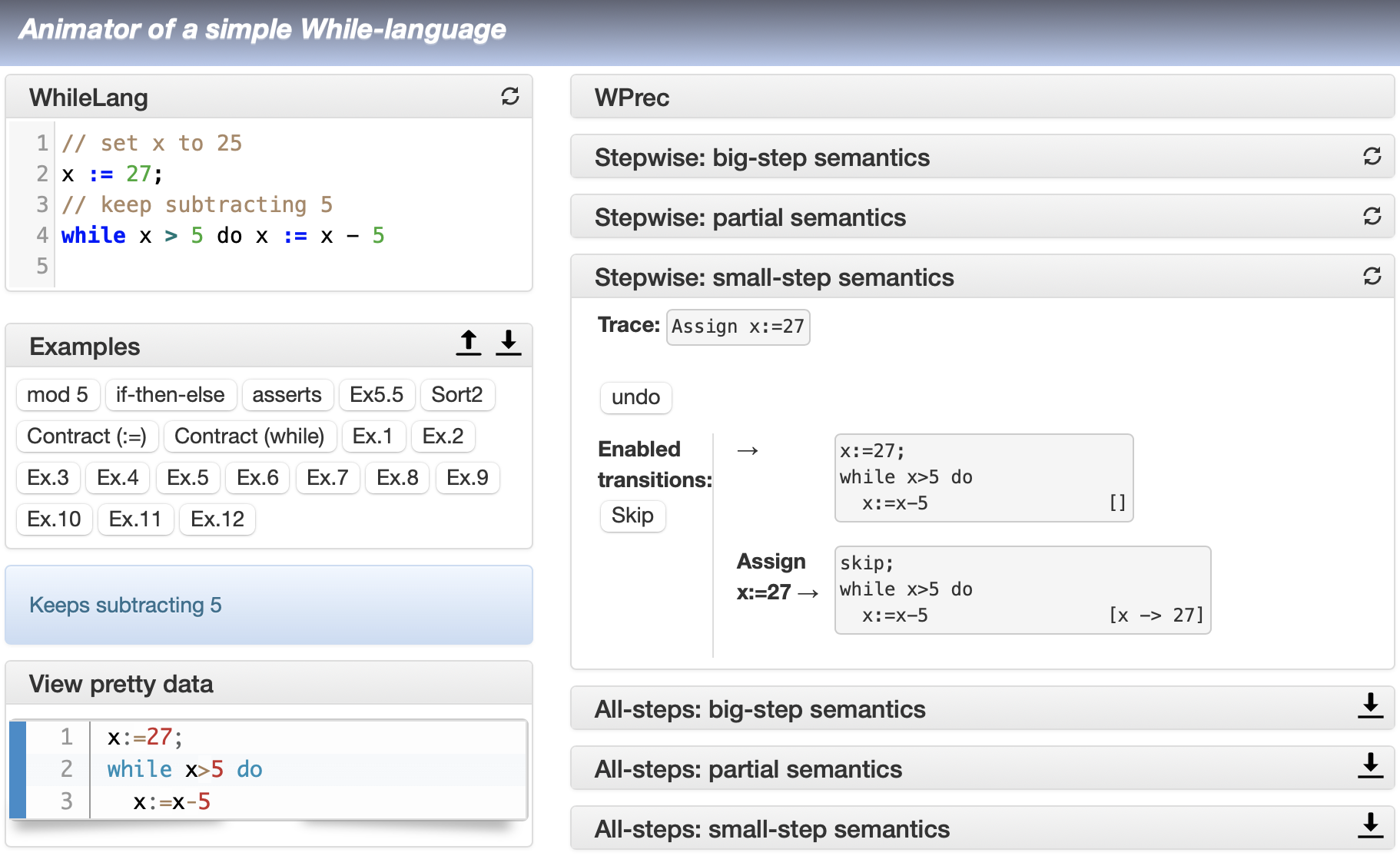}
  \caption{Screenshot of the web interface to analyse a simple while-language, available at \surl{https://cister-labs.github.io/whilelang-scala/}}
  \label{fig:while}
\end{figure}

The \CAOS toolset is based on ReoLive,\footnote{\surl{https://github.com/ReoLanguage/ReoLive}} which was developed as an online set of Scala \& JavaScript (JS) tools to analyse Reo connectors~\cite{reolive}.
Currently it also hosts many extensions unrelated to Reo~\cite{goncharov-implementing-ictac-20,cledou-hubs-2021}, where common code blocks can be compiled both to JS (client) and to Java binaries (server), allowing computations to be delegated to a remote server.
Consequently, it became a \textbf{monolithic} implementation with many \textbf{replicated} blocks of code for different independent extensions, and it is non-trivial to \textbf{reuse} it for different projects.
Our alternative \CAOS toolset aims at addressing these issues,
targeting the following requirements:
\begin{itemize}
  \item \textbf{\textsf{R1}:} \CAOS should use a general programming language, facilitating adoption and supporting more complex back-ends when desired;
  \item \textbf{\textsf{R2}:} The output from \CAOS-supported implementation should be easy to execute and use, without requiring specific platforms or complex installations;
  \item \textbf{\textsf{R3}:} \CAOS should be easily reused, and \CAOS-supported implementations should be modular and easily extended with new analyses.
\end{itemize}
Guided by these requirements, our \CAOS toolset is implemented in Scala (\textsf{R1}), compiles to JS that generates intuitive and interactive websites (\textsf{R2}), and includes a simple-to-extend API that facilitates its usage and reuse by other developers (\textsf{R3}).
By using the \CAOS toolset, one can produce a webpage such as the one in \cref{fig:while}. This webpage has an input text box and a collection of widgets that depict or animate different analyses over the input program, exploiting possible operational semantics when applicable.
This example will be further detailed in \cref{sect:while}, which analyses
a simple while-language (with contracts).
 
\CAOS includes dedicated support for SOS, by animating, depicting, or comparing terms that implement a \texttt{next} and an \texttt{accepting} method.
It further supports building SOS for networks of interacting components, mentioned in \cref{sect:realisability}.
Similar approaches to support the development of language semantics exist, such as the ones below, which do not address all of the 3 requirements above.
 
\smallskip
The \textbf{Maude language} and toolset~\cite{DBLP:conf/rta/ClavelDELMMT03} focus on how to specify (1) a configuration (a state) using a sequence of characters, and (2) a set of possible rewrite rules capturing how configurations can be modified.
It further provides a set of constructs to facilitate the creation of new syntactical notations, such as marking operators as being associative and with a given identity. Maude includes well polished model checkers and other analysis tools; other model checkers (e.g., mCRL2~\cite{BGKLNVWWW19}, UPPAAL~\cite{David2015}) also have specification languages with an operational semantics, restricted by design to provide better model-checking support.
These approaches provide a similar functionality but do not target our requirements.

\smallskip
\textbf{Racket} (and its DrRacket graphical frontend)~\cite{DBLP:journals/cacm/Flatt12} is a \emph{Language-Oriented Programming Language}, i.e., a language meant for making languages. It is widely adopted and comes with a large collection of libraries, and includes a set of constructs that facilitate the creation of new syntactical notations, bundled as new languages, allowing multiple languages in a program to exist and to be created on the fly. Embedded in Racket, PLT Redex~\cite{DBLP:books/daglib/0023092}
is a domain specific language for specifying and debugging operational semantics, which receives a grammar and reduction rules and supports an interactive exploration of the terms. Arguably, Racket is a general purpose language (\textsf{R1}), although less adopted than Java or Scala, with extension mechanisms to support reusability (\textsf{R3}), 
and which we believe to be harder to deploy products (\textsf{R2}).
\smallskip
Some \textbf{teaching languages}, such as Pyret~\cite{pyret19}, are designed to be compiled to JavaScript and to be used when teaching introductory computing, balancing expressiveness and performance. It includes a powerful runtime to hide from the user some of the intricacies and limitations of JS. These languages often include visualisation libraries and engines that engage their target audience: students learning programming concepts.
These languages do not share the precise same functional goal, and do not use a general programming language (\textsf{R1}), but can often produce easy-to-run code (\textsf{R2}) and be extendable (\textsf{R3}).

\smallskip
\CAOS is particularly useful for users familiarised with Scala/Java, and less to users with some experience in languages such as Racket or Pyret.
\smallskip
\noindent\textbf{Paper structure:}
This paper starts by describing our experience with \CAOS both in
a teaching (\cref{sect:while}) and
a research (\cref{sect:realisability})
context, focused on what can be produced using the toolset. \Cref{sect:framework} describes how the \CAOS toolset is structured and how it can be used by others, and \Cref{sect:conclusions} concludes this paper.
\Cref{sect:lambda} provides a step-by-step tutorial how to use the \CAOS toolset.

\section{Use-case: a While-language for Teaching}
\label{sect:while}
In the context of a university course, students were taught about natural and operational semantics, and how to infer weakest preconditions. We, as teachers, used a simple while-language with integers to describe these concepts.
We created a simple website \textbf{in a couple of days} using \CAOS, depicted in \cref{fig:while},
improving core widgets over the period of one week.
Note that we were familiarised with the tools and
had some experience with writing parsers in Scala.
This website was
used by the students to experiment and gain better insights over the concepts.

\cref{fig:while} illustrates the compiled output of \CAOS: a collection of widgets that always includes an input widget (here called \code{WhileLang}) and a list of example input programs. The other widgets are custom-made, and include: (1) \emph{visualisation} of a string produced from the program, representing plain text, code, or a mermaid diagram (a popular Markdown-like language for diagrams);
\footnote{\surl{https://mermaid-js.github.io/mermaid}} and (2) \emph{execution} given a \scala{next} function that evolves the program, which can be presented either step-wise (interactive)
or as a single state diagram with all reachable states.
\CAOS also provides widgets for 
(3) \emph{comparing} two program behaviours using
bisimilarity or trace equivalence; and
(4) \emph{checking} for errors or warnings in a program.

\Cref{fig:while} depicts a visualisation of the source code (bottom left) and a step-wise evolution using a small-step semantics with a textual representation (right), and the remaining widgets are collapsed. These collapsed widgets use different semantics, provide a view of all steps, or calculate the weakest preconditions, and are not processed while collapsed.
Students could use better understand which rules could be applied at each moment, and navigate through the state space.
\section{Use-case: Analysing Choreographies in Research}
\label{sect:realisability}

\CAOS can be used to illustrate research concepts using prototyping tools.
We used it, for example, when investigating choreographic languages.
A choreographic language describes possible sequences of interactions between agents, e.g.,

\smallskip
{\centering \cod[Work,Done]{ctr->wrk1:Work;ctr->wrk2:Work;(wrk1->ctr:Done||wrk2->ctr:Done)}
\smallskip

}
\noindent
captures a scenario where a controller \cod{ctr} delegates some \cod[Work]{Work} to two workers, and they reply once they are \cod[Done]{Done}.
Together with Guillermina Cledou and Sung-Shik Jongmans we published several choreography analyses supported by \CAOS-based prototypes, investigating how to detect that the behaviour of the local agents induce the global behaviour (known as realisability) using a novel underlying mathematical structure~\cite{Edixhoven-branching-2022,pomset-realisability22} (\surl{https://lmf.di.uminho.pt/b-pomset})
and how to generate APIs that statically guarantee that the local agents follow their interaction protocol~\cite{cledou-apigeneration-2022,jongmans-st4mp-2022} (\surl{https://lmf.di.uminho.pt/{pompset,st4mp}}).
\begin{figure}[t]
  \centering
  \begin{minipage}[t]{0.38\textwidth}
    \wrap{\includegraphics[width=\textwidth]{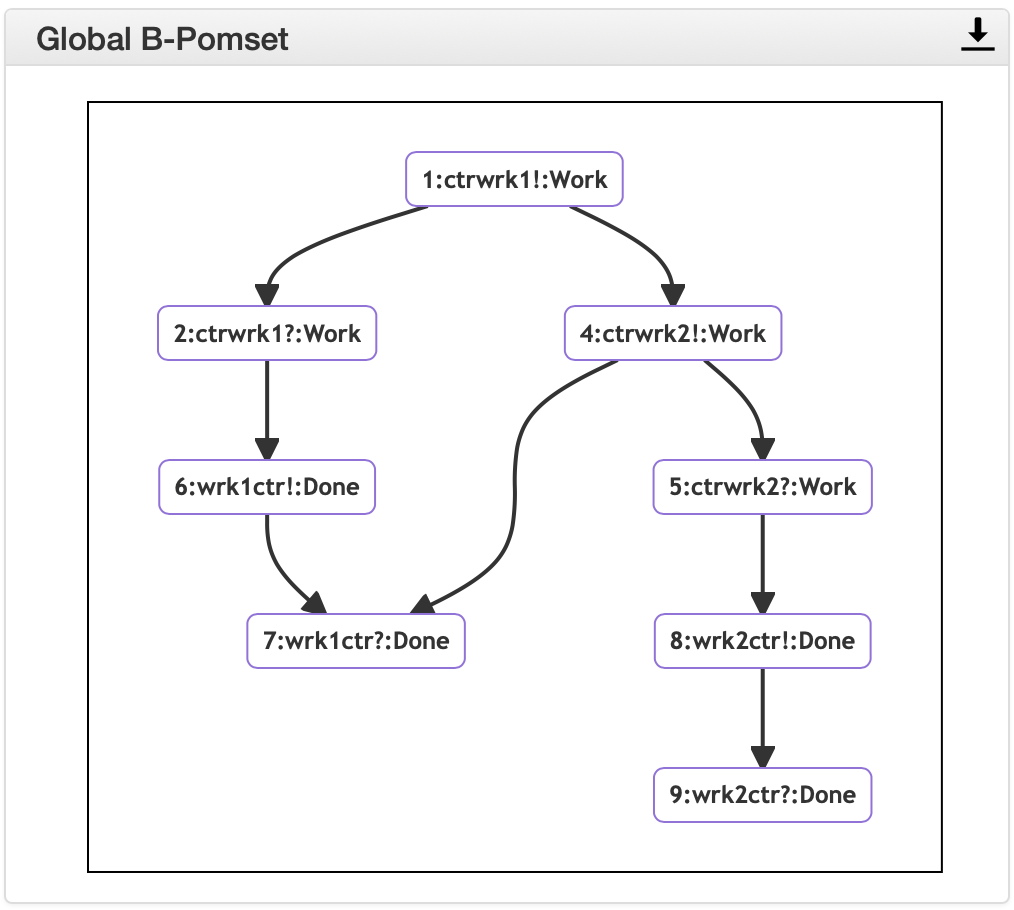}\\
    \includegraphics[trim = 0mm 106mm 0mm 0mm,clip,width=\textwidth]{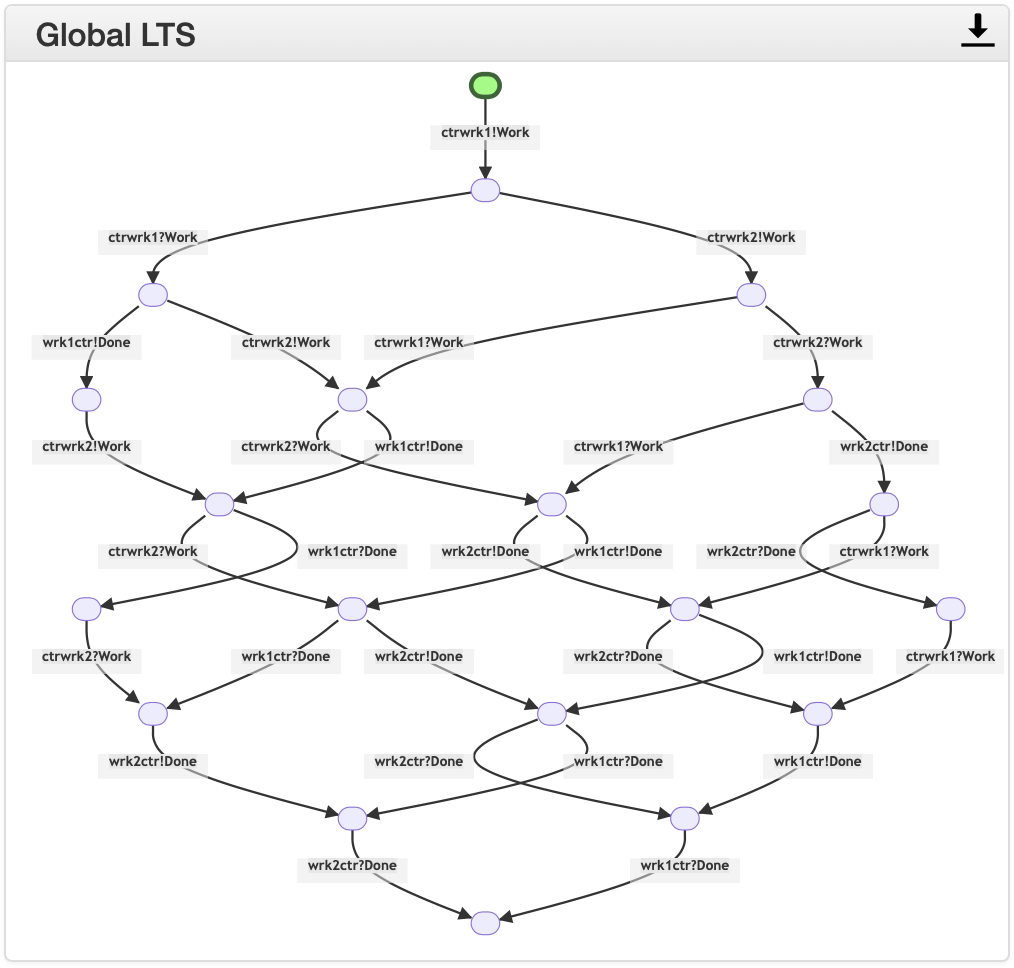}}
      \end{minipage}
  \begin{minipage}[t]{0.61\textwidth}
    \wrap{\includegraphics[trim = 0mm 43mm 0mm 0mm,clip,width=\textwidth]{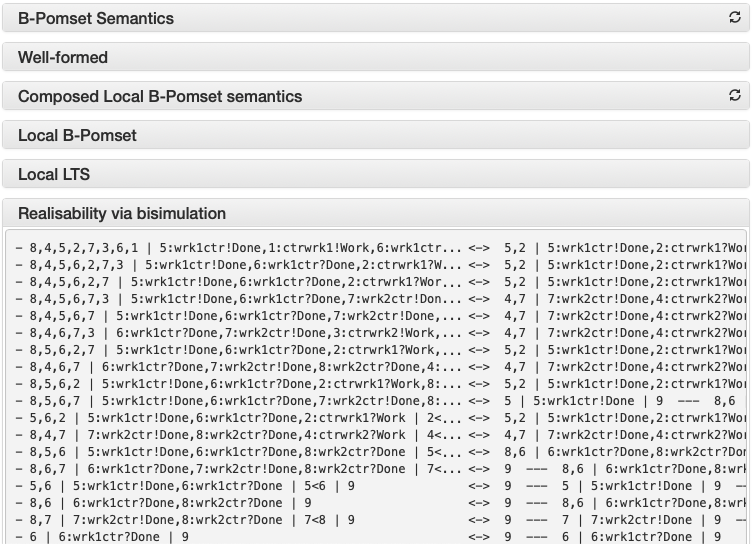}\\
        \includegraphics[trim = 0mm 96mm 0mm 0mm,clip,width=\textwidth]{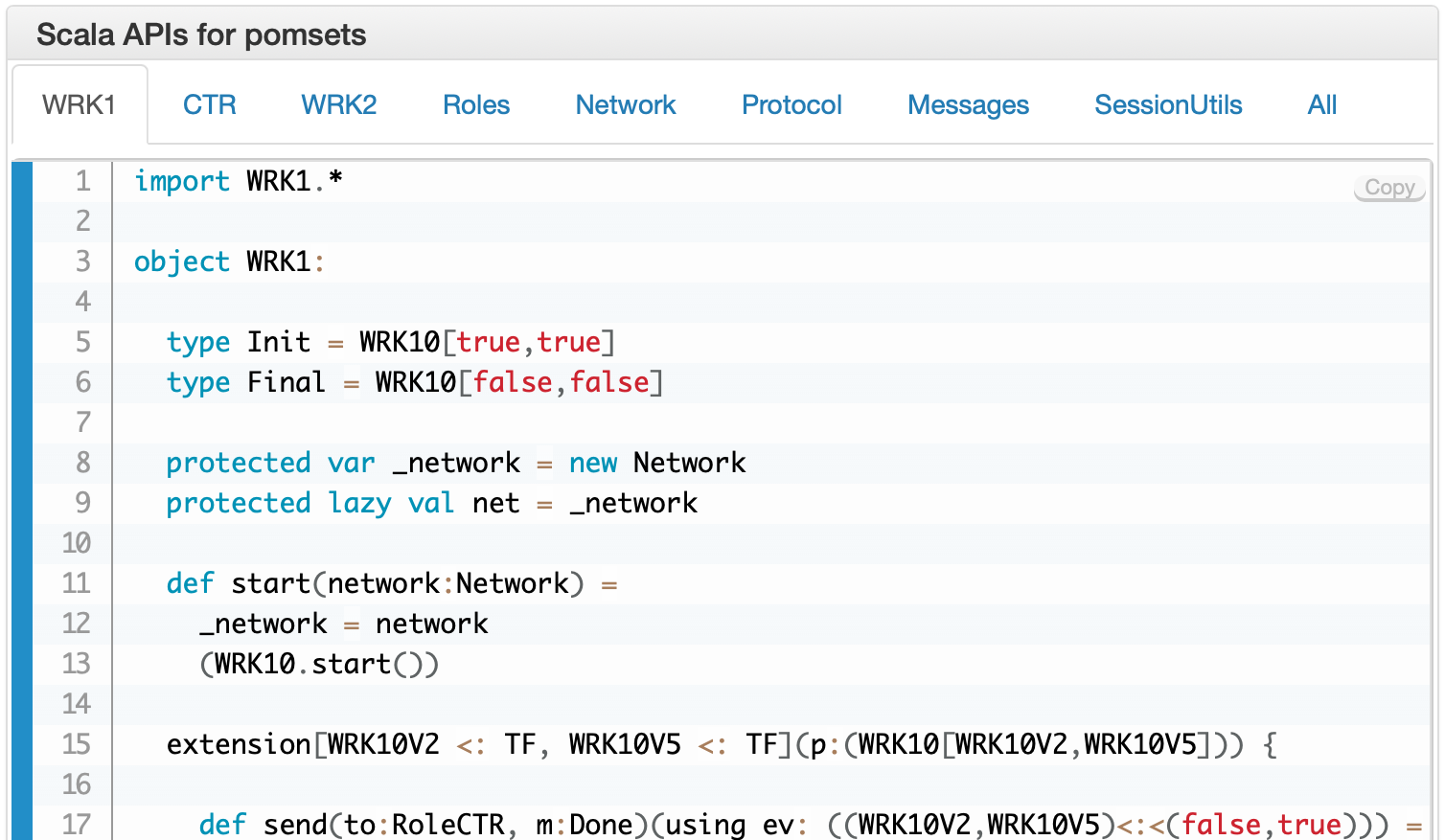}}
  \end{minipage}
  \caption{Analysis of branching pomsets produced by \CAOS from a choreography language}
  \label{fig:pomsets}
\end{figure}

An underlying mathematical structure was used to give semantics to choreographies: branching pomsets~\cite{Edixhoven-branching-2022} (which are similar to event structures~\cite{DBLP:journals/tcs/NielsenPW81,DBLP:conf/birthday/CastellaniDG19}). As shown in \cref{fig:pomsets}, using \CAOS it was possible to:
  (1) \emph{visualize} the pomset structure (top left);
    (2) \emph{execute} a pomset ({\small\sf B-Pomset Semantics})
    and the composition of its projections to each agent involved ({\small\sf Composed Local B-Pomset Semantics});
  (3) \emph{check well-formedness} ({\small\sf Well-formed}), a novel syntactic (sound but incomplete) realisability check;
    (4) \emph{check realisability} using a (complete but more complex) search for a bisimulation between the global behaviour and the composed behaviour of the projections ({\small\sf Realisability via bisimulation}); and
      (5) \emph{generate Scala code} with libraries that can guarantee at compile time that local agents obey the expected protocol   (bottom right).
\CAOS provides constructors for the composition of the behaviour of the local agents and for the search for bisimulations.
 
Setting up each of these websites \textbf{took around half a week} of work by one person.
During our investigation, the \CAOS-supported implementation was a \emph{crucial} mechanism to 
experiment with many variations of
the semantics and projections of the choreography language, of the pomset structure, and of the realisability analysis, ultimately converging to the current version.
\section{\CAOS framework} 
\label{sect:framework}

This section describes what \CAOS provides and how to use it to produce animators such as the ones in \cref{sect:while,sect:realisability}. A step-by-step guide on how to use \CAOS is included in \cref{sect:lambda}, using a lambda-calculus-based language to illustrate~it.
\begin{figure}
  \begin{tikzpicture}[>=stealth',thick]
    \tikzstyle{doc}=[shape=document,thick,draw=black,minimum height=6mm,
                      font=\sf\smaller,align=center]
    \tikzstyle{user}=[fill=blue!20]
    \tikzstyle{lbl}=[font=\scriptsize,align=center,fill=none,above,
                     fill opacity=0.8,text opacity=1,inner sep=3pt]
    \node[doc,user](analysis){Analsysis.scala};
    \node[doc,user,right=1.5 of analysis](config){Configuration.scala};
    \node[doc,above=0.5 of analysis](caos){\CAOS.scala};
    \node[doc,right=1.5 of config](compiled){compiled.js};
    \node[doc,right=1.5 of compiled](site){index.html};

    \draw[->] (config) to node[lbl]{imports}(analysis);
    \draw[<-,rounded corners] (caos) -| node[lbl,right,pos=0.8]{extends}(config);
    \draw[<-] (caos) to node[lbl,right]{imports}(analysis);
    \draw[->] (config) to node[lbl]{compiles}(compiled);
    \draw[<-] (compiled) to node[lbl]{inlcudes}(site);
  \end{tikzpicture}
  \caption{Architecture of a Scala project that uses \CAOS}
  \label{fig:caos-architecture}
\end{figure}
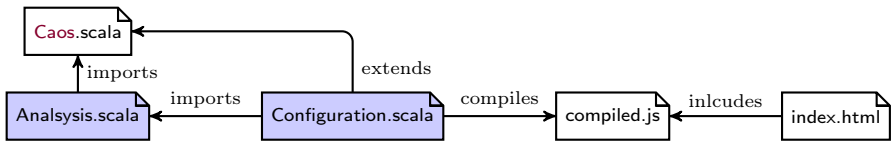

\Cref{fig:caos-architecture} depicts the structure of a typical Scala project that uses \CAOS.
The user
provides data structures for the input language together with functions to parse this language and to compute analysis (\textsf{Analysis.scala}), and
compiles a collection of widgets that use these functions using the provided constructors (\textsf{Configuration.scala}).
Compiling this configuration yields a JS file that is called by a provided HTML file.
The \textsf{Configuration} is an object that extends an associated class in \CAOS and holds: 
 the \textbf{name} of the language and the website;
 the \textbf{parser} for the language;
 a list of \textbf{examples}, each as a triple (name, program, description); and
 a list of \textbf{widgets} using the provided constructors described in \cref{sect:configuring}.
The compilation is configured in a dedicated \textsf{build.sbt} file that selects the ScalaJS compiler, the main file that will be compiled, and location of the final JS file.
The user can specify operational rules by extending
a \scalac{SOS[Act,State]} class providing a function \scalac{next(s:State): Set[(Act,State)]} that, given a \scalac{State} \scalac{s}, returns a set of new states labelled by an \scalac{Act}ion.
These instances can be animated, compared, or combined using provided widget constructors. For example, \scala{lts(makeBPomset,BPomsetSOS,bp=>"")} builds the \scala{Global LTS}
in \cref{fig:pomsets}, where \scala{makeBPomset} creates the pomset structure, \scala{BPomsetSOS} extends the \scalac{SOS} class, and \scala{bp=>""} hides state names.
\section{Conclusions and lessons learned}
\label{sect:conclusions}

This paper follows a \emph{computer-aided design approach for formal methods} by means of \CAOS, introducing its associated toolset and sharing experiences of its application to develop operational semantics of different systems.
During the development of new structures and operational semantics, the \CAOS toolset provided support to quickly view, simulate, and compare different design choices.

We were able to efficiently identify possible problems and solutions, with a small investment of time in tool development. We further claim that the \CAOS toolset is reusable, provides intuitive outputs, and is expressive by using a general programming language. By using standard HTML and CSS, the resulting websites are also easily customisable if desired.

Currently we are considering two possible improvements. On one hand,
supporting a lightweight server
(inspired in ReoLive~\cite{reolive} but using, e.g., \surl{https://http4s.org}) that could be used to delegate heavier tasks, such as the usage of a model-checker. On the other hand, to better support the parser development with tools such as \surl{https://antlr.org} that support both JS and Java backends, instead of using parsing combinators.
All tools are available as open-source, and we welcome any feedback, contribution, or sharing of experiences.

\bibliographystyle{splncs04}
\bibliography{src/bib}
\clearpage
\appendix

\section{Tool Demonstration: Animating the iLambda language}\label{sect:lambda}

\begin{figure}
  \centering
  \includegraphics[width=\textwidth]{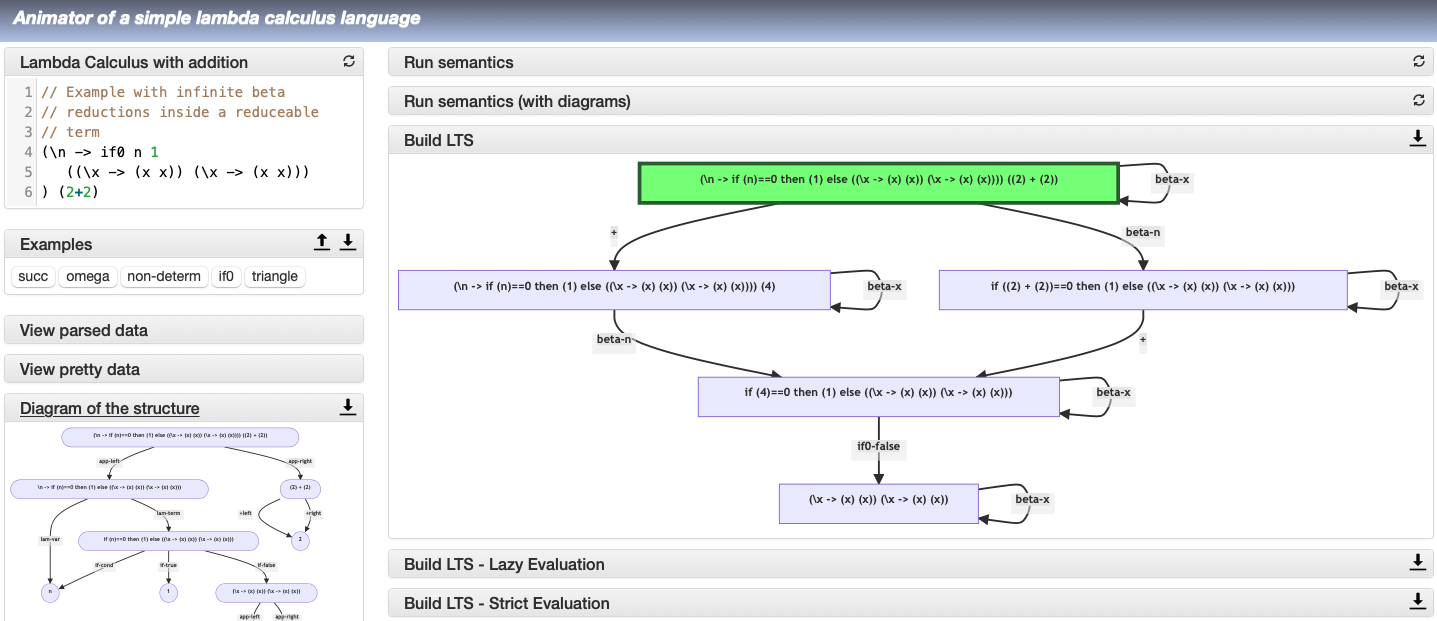}
  \caption{Generated site for the \filename{iLambda} language}
  \label{fig:iLambda}
\end{figure}

This appendix demonstrates how to incorporate and use \CAOS in the context of a lambda-calculus language with integers---which we call \filename{iLambda}---using different evaluation strategies. A video demoing this example can be found at \surl{https://zenodo.org/record/7876060}. The resulting webpage is depicted in \cref{fig:iLambda} and can be found online:
\begin{itemize}
  \item \textbf{Full source-code:} \surl{https://github.com/arcalab/lambda-caos}
  \item \textbf{Generated site:} \surl{http://lmf.di.uminho.pt/lambda-caos/}
\end{itemize}
This demonstration consists of a step-by-step guide explaining how to replicate the creation of this \filename{iLambda} Scala project.

\subsection{Requirements}
You will be programming in Scala and compiling the code to JavaScript (JS), but you do not need to install neither Scala nor a JS engine. Instead, you will need:
\begin{itemize}
  \item JVM (1.8+): Java Virtual Machine (there is no need for a Java compiler);
  \item SBT: Builder tool for Scala projects, similar to Maeven for Java;
  \item An internet browser with JS support, such as Chrome or Firefox;
  \item Your favourite Scala editor (a text editor is enough).
  \end{itemize}
\subsection{Bootstrapping and Compiling}
You will create a file structure recognised by SBT, download \CAOS, and edit two configuration files for SBT. We will call our demo project \filename{iLambda}, and will structure the files as follows:
\begin{itemize}
  \item \filename{build.sbt} -- the main configuration file of the project;
  \item \filename{project/plugins.sbt} -- where we include the ScalaJS plug-in to compile into JS;
  \item \filename{lib/\CAOS} -- where we include all \CAOS files, as-is in its git repository;
  \item \filename{src/main/scala/iLambda} -- where we include all the source-code of our project.
  \item \filename{src/main/scala/iLambda/frontend/Main.scala} -- where we include the \filename{main} method, which we will configure as the starting point of the JS compilation.
\end{itemize}
If you are using git, you can keep track of new versions of \CAOS by importing it as a git submodule, e.g., with the command \bash{git submodule add <link-to-\caos{}\-local-git-repo> lib/\caos{}}.
Assuming that the \filename{Main.scala} file mentioned before has a \filename{Main} object with a \filename{main} method, you now need to edit the two \filename{.sbt} files as in \cref{fig:sbt}.
\begin{figure}
\centering
\begin{tikzpicture}
  \node[shape=document,thick,draw=black,minimum height=56mm,
        font=\tt\smaller\smaller,align=left](conf) {
\bl{val} \pr{\caos} = project.in(file(\gr{"lib/\caos{}"}))
\\~~.enablePlugins(ScalaJSPlugin)
\\~~.settings(scalaVersion := \gr{"3.1.1"})
\\
\\\bl{val} \pr{iLambda} = project.in(file(\gr{"."}))
\\~~.enablePlugins(ScalaJSPlugin)
\\~~.settings(
\\~~~~name := \gr{"iLambda"},
\\~~~~version := \gr{"0.1.0"},
\\~~~~scalaVersion := \gr{"3.1.1"},
\\~~~~scalaJSUseMainModuleInitializer := true,
\\~~~~\ext{Compile / mainClass} := Some(\gr{"iLambda.frontend.Main"}),
\\~~~~\ext{Compile / fastLinkJS / scalaJSLinkerOutputDirectory} :=
\\~~~~~~baseDirectory.value / \gr{"lib"} / \gr{"\caos{}"}/
\\~~~~~~\gr{"tool"} / \gr{"js"} / \gr{"gen"},
\\~~~~libraryDependencies += \gr{"org.typelevel"} \\\~~~~~~\gr{"cats-parse"} \\\~~)
\\~~.dependsOn(\pr{\caos})
};

  \node[shape=document,thick,draw=black,minimum height=56mm,
        font=\tt\smaller\smaller,align=left,right=1 of conf](plugin) {
        addSbtPlugin(
      \\~~~\gr{"org.scala-js"} \      \\~~~\gr{"sbt-scalajs"} \      \\~~~\gr{"1.7.1"}
      \\)\\~\\~\\~\\~\\~\\~\\~\\~\\~\\~\\~\\~\\~\\~
  };

  \node[font=\sf\smaller,below]at(conf.south) (scalatxt) {build.sbt}; 
  \node[font=\sf\smaller,below]at(plugin.south) (pi) {project/plugin.sbt}; 
\end{tikzpicture}
\caption{Build files for SBT to configure the compilation of the project into JavaScript}
\label{fig:sbt}
\end{figure}

To compile the frontend one just needs to run in the command line \bash{sbt fastLinkJS} at the root, in the same folder as \filename{build.sbt}. The main method in \filename{src/main/scala/iLambda/frontend/Main.scala} is then compiled into the file 
\filename{lib/\caos{}/tool/js/gen/main.js}.

\subsection{Developing iLambda}
Before populating the \filename{Main.scala} file with the website information, we need to define a set of core functions and data types. In this case we developed the following classes:
\begin{itemize}
  \item \filename{iLambda/syntax/Program.scala} -- where we defined the structure of our abstract syntax tree, in this case: 
    \\{\footnotesize\tt
  $~$~~\bl{enum} \pr{Term}:\\
  $~$~~~~\bl{case} Var(x:\pr{String})\\
  $~$~~~~\bl{case} App(e1:\pr{Term}, e2:\pr{Term})\\
  $~$~~~~\bl{case} Lam(x:\pr{String}, e:\pr{Term})\\
  $~$~~~~\bl{case} Val(n:\pr{Int})\\
  $~$~~~~\bl{case} Add(e1:\pr{Term}, e2:\pr{Term})\\
  $~$~~~~\bl{case} If0(e1:\pr{Term}, e2:\pr{Term}, e3:\pr{Term})
  }
  \\[-2mm]
  \item \filename{iLambda/syntax/Parser.scala} -- where we define the parser, i.e., a function \filename{Parser.parseProgram} that converts a string into a \filename{Program.Term} defined above.
  \\[-2mm]
  \item \filename{iLambda/syntax/Show.scala} -- where we define functions to visualise terms as strings, including strings that denote Mermaid diagrams.
  \\[-2mm]
  \item \filename{iLambda/backend/Semantics.scala} -- where we extend the \filename{\caos.sos.SOS} class by implementing the \filename{next} function with different ways to rewrite a \filename{Term}, e.g., by applying a lambda abstraction to another term (called beta reduction).
    \\[-2mm]
  \item \filename{iLambda/backend/\{LazySemantics,StrictSemantics\}.scala} -- where we implement a variation of the previous semantics, restricting only to rewrite rules with a specific evaluation order; the Lazy semantics gives priority to the beta reduction, while the Strict semantics evaluates the arguments first.
\end{itemize}
\subsection{Configuring \CAOS} \label{sect:configuring}

As a final step, we now populate the \filename{Main.scala} with a configuration that produces the widgets depicted in \cref{fig:iLambda} that analyse lambda terms. A possible definition of this class can be found below, followed by a description of the core widget constructors.

\begin{lstlisting}[style=scalastyle,morekeywords={[2]private}]
def main(args: Array[String]):Unit =
  $\caos$.frontend.Site.initSite[PTerm](MyConfig)

object MyConfig extends Configurator[Term]:
  val name = "Animator of a simple lambda calculus language"
  override val languageName: String = "Lambda Calculus with addition"
  
  val parser = iLambda.syntax.Parser.parseProgram
  
  val examples = List(
    "succ" -> "(\x -> x + 1) 2" -> "Adds 1 to number 2",
    ...)
  
  val widgets = List(
    "View parsed data"             -> view(_.toString, Text),
    "View pretty data"             -> view(Show(_), Code("haskell")),
    "Diagram of the structure"     -> view(Show.mermaid, Mermaid),
    "Run semantics" -> steps(e=>e, Semantics, Show(_), Text),
    "Run semantics (with diagrams)"->
                       steps(e=>e, Semantics, Show.mermaid, Mermaid),
    "Build LTS"                    -> lts(e=>e, Semantics, Show(_)),
    "Build LTS - Lazy Evaluation"  -> lts(e=>e, LazySemantics, Show(_)),
    "Build LTS - Strict Evaluation"-> lts(e=>e, StrictSemantics, Show(_))
    "Find bisimulation: given 'A B', check if 'A ~ B'" ->
      compareBranchBisim(Semantics, Semantics,
                         getApp(_).e1, getApp(_).e2, Show(_), Show(_)),
  )

  private def getApp(t:Term): Term.App = t match
    case a:Term.App => a
    case _ => sys.error("Input must be an application \"A B\" to compare \"A\" and \"B\".")
\end{lstlisting}

Some of the most used constructors of widgets are summarised below, properly documented in the \CAOS (open) source-code.
\begin{itemize}
  \item \scalac{view[Pr](v: Pr=>String, t:ViewType)}
  \\visualises the input program \scalac{p} represented by \scalac{v(p)} as text, code, or a mermaid diagram, depending on \scalac{t};
  \item \scalac{steps[Pr,A,St](initSt:Pr=>St, sos:SOS[A,St], v:St=>String, t:ViewType)}
  \\depicts the evolution of the state \scalac{initSt(p)} using the \scalac{sos} semantics, and visualising the state using \scalac{v(p)} and \scalac{t} as before;
  \item \scalac{lts[Pr,A,St](initSt:Pr=>St, sos:SOS[A,St], vs:St=>String,va:A=>String)}
  \\similar to before, but unfolds the \scalac{sos} semantics as many times as possible, producing a diagram with the reachable state space;
  \item \scalac{compareBranchBisim[Pr,A,S1,S2](init1:Pr=>S1,init2:Pr=>S2}
  \\    \scalac{$\hspace*{47.5mm}$,sos1:SOS[A,S1],sos2:SOS[A,S2]}
  \\    \scalac{$\hspace*{47.5mm}$,v1:S1=>String,v2:S2=>String)}
  \\returns either a branching bisimulation between \scalac{init1(p)} (using \scalac{sos1}) and \scalac{init2(p)} (using \scalac{sos2}), an explanation why no bisimulation was found, or a timeout message, using \scala{v1} and \scala{v2} to visualise the states;
  \item \scalac{check[Pr](a: Pr=>Seq[String])}
    produces a set of warning messages \scalac{check(a)}, possibly throwing exceptions that are caught by the frontend, and remains invisible if no messages are produced.
\end{itemize}
\subsection{Defining SOS semantics}
Recall that several widgets receive an \scalac{SOS[A,St]} object with actions of type \scalac{A} and states of type \scalac{St}. These interfaces are defined by \CAOS; for example, part of the lazy-semantics of iLambda is implemented below, and does not exploits the notion of acceptance.

\begin{lstlisting}[style=scalastyle]
object LazySemantics extends SOS[String,Term] {
  /** What are the set of possible evolutions (label and new state) */
  def next[A>:String](t: Term): Set[(A, Term)] = t match {
    // Cannot evolve variables
    case Var(_) => Set()
    // Evolve body of a lambda abstraction
    case Lam(x, e) =>
      for (by, to) <- next(e) yield by -> Lam(x, to)  
    // Apply a lambda abstraction
    case App(Lam(x,e1),e2) => Set(s"beta-$\texttt{\textcolor{webgreen}{\$}}$x" -> Semantics.subst(e1,x,e2))
    // Try to evolve the left of an application first
    case App(e1, e2) =>
      next(e1).headOption match 
        case Some(head) => Set(head._1 -> App(head._2,e2))
        case None =>       for (by,to) <- next(e2) yield by -> App(e1,to)
    // Remaning cases...
}}
\end{lstlisting}

To help analysing \textbf{concurrent systems}, \CAOS further provides a constructor to build an \scalac{SOS} of a network of local states. Note that our iLambda example does not use this constructor. More specifically, one can use the function \scala{caos.sos.Network.sos(sync, relabel, localSOS)} that builds a new \scalac{SOS} given:
\begin{itemize}
  \item \scala{sync} -- a function that checks which combination of labels from individual participants can be taken in the same step;
  \item \scala{relabel} -- a function that renames a valid combination of labels into a new (global) label;
  \item \scala{localSOS} -- an \scalac{SOS} object for the behaviour of each local participant. 
\end{itemize}
These functions use a notion of state that consists of list of local states and an object that captures the state of the network. This was used, for example in the analysis of realisability of branching pomsets~\cite{Edixhoven-branching-2022,pomset-realisability22}.
\subsection{Wrap up}

This \code{iLambda} is a simple example, available online, and produced only to illustrate how to bootstrap a project that uses \CAOS. Most of this project was developed in \textbf{around 1 day}, in part due to the familiarity of the developers with these libraries and with the usage of parsing combinators. We demonstrate here only a small subset of the available widgets, focused on the most used ones; more documentation and examples can be found online.
\end{document}